\documentstyle[12pt]{article}
\input {epsf}
\def\beq{\begin{equation}}
\def\eeq{\end{equation}}
\def\beqn{\begin{eqnarray}}
\def\eeqn{\end{eqnarray}}

\def\ts{\textstyle}
\def\eq#1{eq.~(\ref{#1})}

\def\tr{\mathop{\rm tr}}
\def\Tr{\mathop{\rm Tr}}
\def\VEV#1{\left\langle #1\right\rangle}
\def\p{\partial}
\def\d{{\rm d}}
\def\ee{{\rm e}^}
\def\Lm{\Lambda}
\def\hA{\hat{A}}
\def\hB{\hat{B}}

\def\l{\ell}
\newcommand{\hsig}[1]{%
   \widehat{\sigma}_{#1}}
\newcommand{\tK}[1]{%
   \widetilde{K}_{#1}}
\def\ZZ{{\sf Z}\!\!{\sf Z}}
\catcode`\@=11
\def\numberbysection{\@addtoreset{equation}{section}
\def\theequation{\arabic{section}.\arabic{equation}}}
\def\appendix{\setcounter{section}{0}
        \def\thesection{Appendix \Alph{section}}
        \def\theequation{\Alph{section}.\arabic{equation}}}
\begin{document}
\numberbysection
\begin{titlepage}
\begin{flushright}
 {\normalsize 
    hep-th/9702126 \\ 
     OU-HET 257 \\ 
     February 1997 \\}
\end{flushright}
\vfill
\begin{center}
  {\Large \bf
  Boundary operators and touching of loops\\
  in 2d gravity}
\footnote{
 This work is supported in part
 by the Grand-in-Aid for Scientific Research Fund (2690)
 from the Ministry of Education, Science and Culture, Japan.
}
\vfill
 {\bf Masahiro Anazawa}
 \footnote{e-mail: anazawa@funpth.phys.sci.osaka-u.ac.jp
 \quad JSPS research fellow}\\ 
\vfill
 {\it
    Department of Physics,\\
    Graduate School of Science, Osaka University,\\
    Toyonaka, Osaka, 560 Japan\\
  }
\vfill
\end{center}

\begin{abstract}

   We investigate the correlators in unitary minimal
 conformal models coupled to two-dimensional gravity from the two-matrix
 model.
  We show that simple fusion rules for all of the  scaling operators
 exist.
   We demonstrate the role played by the boundary operators and discuss
 its connection to how loops touch each other.
\end{abstract}
\vspace{0.2in}
\end{titlepage}

\begin{center}
\section{Introduction}
\label{introduction}
\end{center}

 The understanding of two-dimensional quantum  gravity
 has experienced  great progress through
 the study of the matrix models.\footnote
 {See for example \cite{DifGZj} for review.}
 The one-matrix model has infinite number of critical
 points which are considered to represent the $(2m+1,2)$
 minimal conformal models coupled to two-dimensional
 gravity.
   The two-matrix model
 has critical points which correspond to the 
 $(m+1,m)$ unitary minimal conformal models
 \cite{Douglas-Proc,Tada,DKK}.
  In this paper we investigate the unitary minimal model $(m+1,m)$
 coupled to two-dimensional gravity from
 the two-matrix model.

    The emergence of the infinite number of  scaling
 operators
 $\sigma_j$
 is one of the most important properties 
 of the matrix models.
    Before coupled to gravity, the minimal model has finite
 number of primary fields.
 After gravitational dressing, however,
 infinite number of scaling operators emerge. 
    This phenomenon can be understood as follows.
    In the Kac table we can divide the primary fields
 $\Phi_{r,s}$ into those which are inside the the minimal
 conformal grid $1\le r\le q-1,\; 1\le s \le p-1$
 and those outside, which correspond to the null states.
   Before dressed by gravity, the fields outside the minimal
 conformal grid decouple \cite{BPZ} from physical correlators.
  After gravitational dressing, 
 they cease to  decouple \cite{Kitazawa,AGBG}
 and become infinite number of  scaling operators.
   The similar phenomenon has been shown in continuum framework.
 Through the examination of the BRST cohomology of the 
 minimal model coupled to Liouville theory,
 infinite physical states were shown
 to exist \cite{LZ-BMP}.
  These states have their counterparts in the matrix models
 as the scaling operators.  Some of the scaling operators
 do not have
 their counterparts in the BRST cohomology, which we will discuss
 later.

   In ordinary $(p,q)$ minimal conformal model
 the primary fields satisfy  certain fusion rules \cite{BPZ};
 three-point function
 $\langle \Phi_{r_1,s_1}\Phi_{r_2,s_2}\Phi_{r_3,s_3}\rangle$
 is non-vanishing only when
\beqn
  1+|r_1-r_2|\le r_3 \le \mbox{\rm mim}
 \{r_1+r_2-1,p \},\;\; r_1+r_2+r_3=\mbox{\rm odd}
 \nonumber \\
  1+|s_1-s_2|\le s_3 \le \mbox{\rm mim}
 \{s_1+s_2-1,q \},\;\; s_1+s_2+s_3=\mbox{\rm odd}
\;.
\eeqn
    It is interesting to examine how the fusion rules change
 when the matter couples to gravity.
   The three-point functions
 involving lower dimensional scaling operators
 were examined
 from the point of view of the generalized
 KdV flow in \cite{DifK} 
 in the case of $(m+1,m)$ unitary matter.
  It was shown that the gravitational primaries 
 $\sigma_j\; (j=1,\cdots,m-1)$
 satisfy fusion rules of BPZ type;
 for $j_1+j_2+j_3\le 2m-1$, 
 $\VEV{\sigma_{j_1}\sigma_{j_2}\sigma_{j_3}}$ is
 non-vanishing only when
\beq
  1+|j_1-j_2|\le j_3 \le j_1+j_2-1
\;.
\eeq
   The fusion rules were also examined in continuum framework
 \cite{AGBG}.
   As for the gravitational descendants, however, we think
 clear results have not been obtained.
    In this paper we would like to clarify the
 fusion rules for all of the scaling operators including
 the gravitational descendants in the case of unitary minimal
 model.

   Macroscopic loop correlators,
 which are the amplitudes of the surfaces with boundaries (loops)
 of fixed lengths, are the fundamental amplitudes of the
 matrix models.
   It was shown \cite{MSS,MS}
 that these correlators have more information
 than those of local operators and that  the latter 
 correlators can be extracted
 from the former  correlators
 explicitly in the case of $c=0,1/2,1$.
   They argued there that macroscopic loops could be replaced
 by a sum of local operators whose wave functions satisfy the
 Wheeler-DeWitt equations.

 We generalize this idea
 to the loop correlators \cite{DKK,AII,AII2,AI}
 in the cases of the general unitary minimal models,
 and  derive the fusion rules for 
 all of the scaling operators.
  First we derive the explicit form of the expansion of
 loops in terms of the scaling operators, 
 and then deduce
 the three-point correlators from the loop correlators
 which were calculated in \cite{AII2,AI} from the
 two-matrix model. 
   We show that the three-point correlators
 of all of the scaling operators satisfy certain simple fusion rules
 and these fusion rules are summarized
 in a compact form as the fusion rules for
 three-loop correlators \cite{AII2}.

  In matrix models, there are infinite subset of the  scaling operators
 $\hsig{j}$ ($j=0$ mod $m+1$)
 which do not have their
 counterparts in the BRST cohomology of the minimal model
 coupled to Liouville theory.
  In the case of one-matrix model, Martinec, Moore and
 Seiberg \cite{MMS} argued that these operators are boundary
 operators which couple to the boundaries of
 two-dimensional surface.
  They proved that one of them is in fact a boundary operator
 which measures the total loop length.
   But little has been discussed on the role played by the rest
 of these operators.
   We examine the geometrical meaning of these operators and
 its connection to the touching of loops
 in the case of general unitary models.
  We show that the boundary operators have the role to
 connect several parts of the loops together.
  We also discuss the relation of the boundary operators
 to the Schwinger-Dyson equations
 proposed in \cite{IIKMNS}.
\begin{center}
\section{Expansion of loops in local operators}
\label{sec:expansion}
\end{center}

 We consider the $(m+1,m)$ unitary minimal model coupled to
 two-dimensional gravity from the two-matrix model
 with symmetric  potential.
 The partition function $Z$ is defined by
\beq
\ee Z =\int \d \hA \d \hB 
~\ee{-\frac N{\Lm} \tr \left( U(\hA)+U(\hB)-\hA\hB \right)}
\;\;,
\eeq
 where $\hA$ and $\hB$ are hermitian matrices, and 
 $U$ is a certain  polynomial.
  In this article, we limit our discussion to the
 two-matrix model with symmetric potential and to
 the critical points which correspond to the unitary
 minimal models.
  In the case of asymmetric potential, some of
 the boundaries (loops) of two-dimensional surface  would have
 fractal dimensions different from the usual dimension
 of length.

  In \cite{MSS}, in the case of the one-matrix model, it was shown
 that the loop operator can be expanded in terms of local
 operators, that is,
 the loop can be replaced with the infinite combination of
 local operators,
  except some special cases.
  It was argued  that this is the case for the
 general $(m+1,m)$ unitary models.
   Whether this replacement can be done safely or not is
 connected with whether the
 corresponding classical solution  exits or not in the limit
 of small length of the corresponding loop.
   This claim is quite natural because
 all of the scaling operators are expressed in term of
 one matrix $\hA$ as 
$
 \sigma_{j}=\Tr(1-\hA)^{j+1/2}=\sum_{n}a_{n}(j) n^{-1}
 \Tr \hA^{n} 
$ in the one-matrix model.

   In the two-matrix model,  it appears that
 this idea is not the case
 since the direct connection of the scaling operators
 to the operators $\Tr\hA^{n}$ or $\Tr\hB^{n'}$ is
 not clear.
   But this expansion is considered to be possible
 by the following reason.
   When one of the loops on two-dimensional surface shrunk
 to a microscopic loop, 
 the loop represents local deformation of the surface.
  The microscopic loop can be
 replaced by the insertions of local operators.
 The loop correlators except one-loop case
 are continuous when the length of one of
 the loops approaches zero, so that we expect that
 a macroscopic loop can also  be replaced by a sum
 of local operators.

  In this section we derive explicitly 
 the expansion of loops in
 local operators in the case of the unitary minimal models.
   Using this relation, we can deduce the amplitudes of
 local operators and those involving both loops and
 local operators from the loop correlators.

 Let us recall that the two-loop correlators in the $(m+1,m)$
 unitary minimal model coupled to two-dimensional gravity are
 \cite{DKK,AII2}
\beqn
 \lefteqn{
 \VEV{w^+(\l_1) w^{\pm}(\l_2)}
 } \nonumber \\
 &&=
 \frac{1}{m}
 \frac{M}{2} \frac{\l_1 \l_2}{\l_1 +(\pm)^m \l_2} \;
 \sum_{k=1}^{m-1} \; (\pm)^{k-1}
 \tK{\frac{k}{m}}(M \ell_1)\;
 \tK{1-\frac{k}{m}}(M \ell_2)\;
 \;\; .
\label{eq:w+w+-}
\eeqn
 for $\l_1\ne\l_2$.
 Here $w^+(\l)$ and $w^-(\l)$ represent loop operators
 which create loops made by
 the matrices $\hA$ and $\hB$ respectively, and
 we introduced a notation
\beq
 \tK{p}(M\l)=\frac{\sin \pi|p|}{\pi/2} K_p(M\l)
\;.
\eeq
 Here the parameter $M$ is defined by
\beq
 \left( \frac M2\right)^2=\frac{\mu}{m+1},\;\;
 \Lm-\Lm_*=-a^2 \mu
 ,
\eeq
 where $\Lm_*$ represents the critical value of the bare
 cosmological constant, and $\mu$ is the renormalized
 cosmological constant.

  The definition of a set of the scaling operators has
 arbitrariness which comes from the contact terms \cite{MSS}.
 As a set of local operators, we take the scaling operators
 $\hsig{j}$
 whose wave functions $\Psi_j(\l)$
 satisfy the (minisuperspace) Wheeler-DeWitt equations
\beq
 \left[ -\left(\l \frac{\p}{\p \l}\right)^2
 +4\mu\l^2 +\left(\frac jm\right)^2 \right]
 \Psi_j(\l) =0
\;.
\eeq
 It was shown \cite{MSS} that 
 these scaling operators correspond to the dressed
 primary fields of the conformal field theory in the
 case of one-matrix model.
   In terms of these scaling operators,
 we can obtain simple fusion rules for three-point
 correlators, which we will show
 in the next section.

 We normalize the wave function of $\hsig{j}$ as
\beq
 \VEV{\hsig{j}\; w^{+}(\l)}
 =\frac jm \left(\frac M2\right)^{\frac jm}
 \;\tK{\frac jm}(M\l) 
 \;\; , \;\;  j\geq 1 \; \ne 0 \;(\mbox{mod}\; m)
 \;\;.
\label{eq:wave-function+}
\eeq
   Note that
 the normalization factor
 $\sin \frac jm \pi$ in $\tK{j}(M\l)$ in
 \eq{eq:wave-function+} is consistent
 because there are no scaling operators
 $\hsig{j}$ for $j=0$ (mod $m$) in the matrix model.
 We can express the right hand side of \eq{eq:w+w+-}
 as an infinite sum in terms
 of $\tK{j}(M\l_2)$ for $\l_1<\l_2$ (see appendix A):
\beqn
 \lefteqn{
 \VEV{w^+(\l_1) w^{\pm}(\l_2)}
 } \nonumber \\
 &&=
 \frac 1m \sum_{k=1}^{m-1} \sum_{n=-\infty}^{\infty} \;
 (\pm)^{k-1}
 \Bigl|\frac km +2n\Bigr|\; I_{|\frac km +2n|}(M\l_1)
 \; \tK{\frac km +2n}(M\l_2)
 \; .
\label{eq:w+w+--exp}
\eeqn
   Comparing \eq{eq:wave-function+} with \eq{eq:w+w+--exp},
 we expect the following expansions of the loop operators
 in term of the local operators:
\beq
``\; w^{\pm}(\l)=\frac 1m \sum_{k=1}^{m-1}
 \sum_{n=-\infty}^{\infty} (\pm)^{k-1}
 \left(\frac M2\right)^{-|\frac km +2n|}
 \; I_{|\frac km +2n|}(M\l)\;\hsig{|k+2mn|}\; "
 \; .
\label{eq:w+--exp}
\eeq
   These expansions are the generalizations of those in the case of
 the one-matrix model \cite{MSS} to the cases of general unitary
 minimal models coupled to two-dimensional gravity.

   Since the loop correlators are symmetric under the interchange
 of two kinds of loops, that is,
 $\VEV{w^+(\l_1)\;w^+(\l_2)}=\VEV{w^-(\l_1)\;w^-(\l_2)}$,
 the wave functions of the scaling operators with respect to
 loop $w^-(\l)$ are read as
\beq
 \VEV{\hsig{j}\;w^-(\l)}=(-1)^{j-1} \VEV{\hsig{j}\;w^+(\l)}
 \;\; .
\label{eq:wave-function-}
\eeq

\begin{center}
\section{Fusion rules for scaling operators}
\label{sec:fusion rules}
\end{center}

 Using the expansion of loops \eq{eq:w+--exp}, we
 can obtain the correlators of the scaling operators
 from loop correlators.\footnote{
 The multi-loop correlators were examined
 in \cite{AI} from two-matrix model.
 These correlators were also examined 
 in \cite{Kostov} from the viewpoint
 of random surfaces immersed in Dynkin diagrams.
 }
  In this section, we show that there are rather simple
 fusion rules for all of the scaling operators.
  The fusion rules involving the gravitational descendants
 ($\sigma_j,\; j\ge m+2$ )
 have not been clear from the point of view of generalized
 KdV flow. 

\subsection{One- and two-point functions}
\hspace{5mm}
 Let us examine one- and two-point functions first.
 Since the one-loop amplitude diverges when the loop length
 approaches to zero, this amplitude include the contribution
 which is not represented by the local operators.
 Extracting the parts proportional to $I_{\nu}$ ($\nu>0$),
 which parts can be considered as the contributions from
 local operators,
  from the one-loop amplitude
\beqn
 \lefteqn{
 \VEV{w^{\pm}(\l)}
 =\left(1+\frac 1m \right)\l^{-1}\left(\frac M2\right)
 \tK{1+\frac 1m}(M\l)
 }
 \nonumber \\
 &&=
 \left(\frac M2\right)^{2+\frac 1m}
 \left( I_{2+\frac 1m}(M\l)-I_{-2-\frac 1m}(M\l)
 -I_{\frac 1m}(M\l)+I_{-\frac 1m}(M\l) \right)
 \; ,
 \nonumber \\
\label{eq:w+-}
\eeqn
 we can obtain the one-point functions of
 the scaling operators
\beq
\label{eq:sig(1)}
 \VEV{\hsig{1}}=-m \left(\frac M2\right)^{2+\frac 2m}
 \;,\;\;
 \VEV{\hsig{1+ 2m}}
 =m \left(\frac M2\right)^{4+\frac 4m} 
 \; ,
\eeq
\beq
\label{eq:sig(j)}
 \VEV{\hsig{j}}=0 \;\; ,\;\; j\ne 1,1+2m
 \; .
\eeq

 Let us turn to the two-point functions.
 Substituting \eq{eq:w+--exp} into \eq{eq:wave-function+},
 we obtain the two-point
 functions
\beq
 \VEV{\hsig{i} \hsig{j} }
 =\delta_{ij}\;j\; \left(\frac M2\right)^{2j/m }
 \;\; ,
 \;\; i,j\ne 0\; (\mbox{mod}\;m)
 \;\; .
\label{eq:sig(i)sig(j)}
\eeq
  Note that we obtain diagonal two-point functions.

\subsection{Three-point functions}
\hspace{5mm} 
 The three-loop correlator from two-matrix model is
 obtained in \cite{AII2} as
\beqn
 \lefteqn{
 \VEV{w^+(\l_1) w^+(\l_2) w^+(\l_3)}
 = -\frac{1}{m(m+1)} \left(\frac M2\right)^{1-\frac 1m}
 \l_1 \l_2 \l_3
 } 
 \nonumber \\
 &&\quad\times 
 \sum_{(k_1-1,k_2-1,k_3-1) \atop \in {\cal D}_{3}^{(m)} } \;
 \tK{1-\frac{k_1}{m}}(M \ell_1)\;
 \tK{1-\frac{k_2}{m}}(M \ell_2)\;
 \tK{1-\frac{k_3}{m}}(M \ell_3)
 \; ,
\label{eq:w+w+w+-2}
\eeqn
  Here  we have denoted by ${\cal D}_{3}^{(m)}$
\beqn
 {\cal D}_3^{(m)}
 &=& \Biggl\{ (a_1,a_2,a_3) \Bigm|
      \sum^3_{i (\not= j)} a_i  - a_j \ge 0 
      \;\mbox{for}\; j=1\sim 3\;,
 \nonumber \\
 &&\quad
       \sum^3_{i=1}a_i = {\rm even}\le 2(m-2)\;,
       a_i=0,1,2,\cdots\;  
       \Biggr\}
\;.
\label{eq:D3(m)}
\eeqn
  It was shown \cite{AII2} that the selection rules
 in eqs.~(\ref{eq:w+w+w+-2}) and (\ref{eq:D3(m)})
 correspond to the fusion rules
 \cite{DifK,AGBG} for the dressed primaries
 ( $\phi_{ii},\; i=1,\cdots,m-1$)
 by studying the small length behavior of the three-loop
 correlator.
  Using the expansion of loop \eq{eq:w+--exp},
 we can show that the selection rules in \eq{eq:w+w+w+-2}
 represent the fusion rules for all of the scaling
 operators in a compact form.
 Let us show this in the following.

 Using the formula,
\beq
 z K_{1-|p|}(z)
 =
 \pi\sum_{n=-\infty}^{\infty}
 \frac{|p+2n|}{\sin \pi|p+2n|} I_{|p+2n|}(z)
 \;\; , 
\label{eq:zK-exp}
\eeq
 we first expand the three-loop correlator \eq{eq:w+w+w+-2}  as
\beqn
 \lefteqn{ \!\!\!
 \VEV{w^+(\l_1)w^+(\l_2)w^+(\l_3)}
 =
 \frac{-1}{m(m+1)}\left(\frac M2\right)^{-2-\frac 1m}\;
 \sum_{{\cal D}_3^{(m)}} \sum_{n_1=-\infty}^{\infty}
 \sum_{n_2=-\infty}^{\infty} \sum_{n_3-\infty}^{\infty}
 } \nonumber \\
 &&
 \ts{
 \left(\frac{k_1}{m}+2n_1\right) \left(\frac{k_2}{m}+2n_2\right)
 \left(\frac{k_3}{m}+2n_3\right) }
 I_{|\frac{k_1}m+2n_1|}(M\l_1) I_{|\frac{k_2}m+2n_2|}(M\l_2)
 I_{|\frac{k_3}m+2n_3|}(M\l_3)
 \; .
 \nonumber \\
\label{eq:w+w+w+-exp}
\eeqn
 Comparing \eq{eq:w+w+w+-exp} with \eq{eq:w+--exp},
 we can extract
 the three-point functions
\beqn
 \lefteqn{
 \VEV{\hsig{|k_1+2mn_1|}\hsig{|k_2+2mn_2|}\hsig{|k_3+2mn_3|} }
 } \nonumber \\
 &&= C_{k_1k_2k_3} \frac{-1}{m(m+1)}
 \prod_{i=1}^{3}(k_i+2mn_i)
 \left(\frac M2\right)^{-2-\frac 1m 
 +\sum_{i=1}^{3}\frac 1m |k_i+2mn_i| }
 \; ,
\label{eq:sigsigsig}
\eeqn
 where
\beq
 C_{k_1k_2k_3}=\left\{
  \begin{array}{rc}
  1 \;\; ,& \;\; (k_1-1,k_2-1,k_3-1)\in{\cal D}_3^{(m)}\\
  0 \;\; ,& \;\;  \mbox{otherwise}
  \end{array}\right.
 \;\; .
\label{eq:Ck1k2k3}
\eeq

   For $n_i=0$, \eq{eq:sigsigsig} is nothing but
 the correlator of the gravitational primaries.
   For the gravitational primaries,
  \eq{eq:sig(i)sig(j)} and \eq{eq:sigsigsig} agree with
 the correlators obtained in
 \cite{DifK} from the generalized KdV flow up to a factor $-2$.
  Note that we obtain, here, the correlators of the gravitational
 descendants as well.
  In \cite{AGBG}, the fusion rules for the gravitational
 primaries were examined in continuum framework.
    We have found here the fusion rules
 for the gravitational descendants
 as well as for the gravitational primaries.
  These fusion rules are similar to those for
 the gravitational primaries
 due to the factor $C_{k_1k_2k_3}$ in \eq{eq:sigsigsig}.

   Introducing the equivalence classes
 $[\hsig{k}]$ by
 the equivalence relation
\beq
 \hsig{k}\sim \hsig{|k+2mn|}\;\;,\;\;
 n\in \ZZ
 \;\;,
\label{eq:sigksim}
\eeq
 we can consider the fusion rules in \eq{eq:sigsigsig}
 as fusion rules
 among $[\hsig{k}]$ ($k=1,\cdots,m-1$).
  Note here that  the class $[\hsig{k}]$ does not
 correspond to
 the set which consist of the gravitational primary
 ${ \cal O}_k$ and its
 gravitational descendants
 $\sigma_l\left({ \cal O}_k\right)\;,l=1,2,\cdots$
 in \cite{DifK} introduced from the viewpoint of KdV flow.

\subsection{Further on the fusion rules}
\hspace{5mm}
    In this subsection, let us examine the fusion rules in \eq{eq:sigsigsig}
 further and summarize the relation of the scaling operators
 to the primary fields in the corresponding
 conformal field theory.

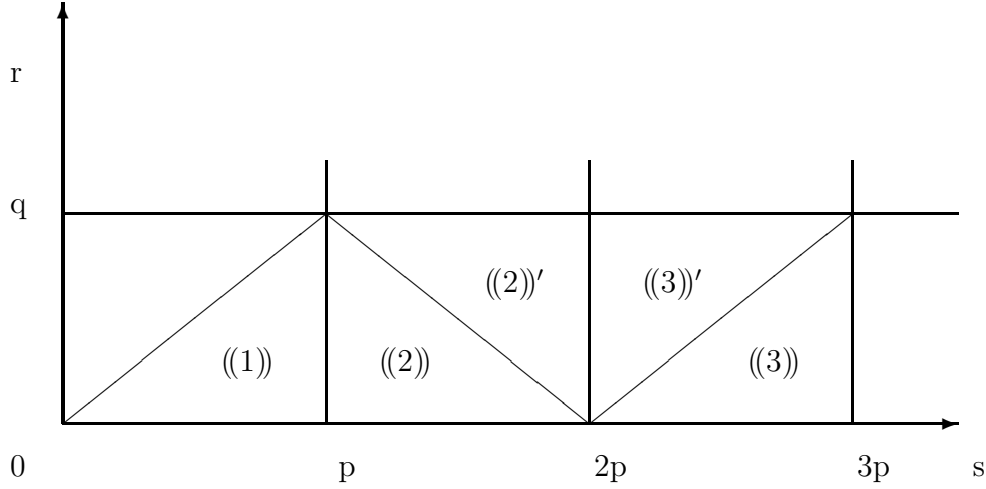
\begin{figure}
 \begin{center}
\setlength{\unitlength}{0.7mm}
\begin{picture}(300,80)(-10,-10)
 \put(0,0){\thicklines\vector(1,0){170} }
 \put(0,0){\thicklines\vector(0,1){80} }
 \put(0,40){\thicklines\line(1,0){170} }
 \multiput(50,0)(50,0){3}{\thicklines\line(0,1){50} }
 \multiput(0,0)(100,0){2}{\line(5,4){50} }
 \put(50,40){\line(5,-4){50} }
 \put(-10,-10){\makebox(8,8)[bl]{0}}
 \put(-10,65){\makebox(8,8)[bl]{r}}
 \put(-10,40){\makebox(8,8)[bl]{q}}
 \put(170,-10){\makebox(8,8)[b]{s}}
 \put(50,-10){\makebox(8,8)[b]{p}}
 \put(100,-10){\makebox(8,8)[b]{2p}}
 \put(150,-10){\makebox(8,8)[b]{3p}}
 \put(30,10){\makebox(10,10)[bl]{$(\!(1)\!)$}}
 \put(60,10){\makebox(10,10)[bl]{$(\!(2)\!)$}}
 \put(130,10){\makebox(10,10)[bl]{$(\!(3)\!)$}}
 \put(80,25){\makebox(10,10)[bl]{$(\!(2)\!)'$}}
 \put(110,25){\makebox(10,10)[bl]{$(\!(3)\!)'$}}
\end{picture}
 \end{center}
\caption{the range of $(r,s)$ }
\label{fig:pic1}
\end{figure}

  In the $(p,q)$ minimal conformal model,
 the primary field $\Phi_{rs}$ has the conformal dimension
\beq
 \Delta_{r,s}=\frac{(pr-qs)^2-(p-q)^2}{4pq}
 \;\; ,
\label{eq:Delta-rs}
\eeq
  where $r$ and $s$ are positive integers. Since we have
$
 \Delta_{r,s}=\Delta_{r+q,s+p}=\Delta_{q-r,p-s}
$,
 the corresponding primary fields can be regarded as the same
 one. The integers $r$ and $s$ can thus be restricted
 in the range
\beq
 1\le r \le q-1 ,\quad
 1\le s \;\ne 0 \;(\mbox{mod}\; p) ,\quad
 pr < qs 
\label{eq:rs-range}
\eeq
 (see fig.~\ref{fig:pic1} ).
   In fig.~\ref{fig:pic1},
 the primary fields in the region $(\!(2)\!)$ or
 $(\!(2)\!)'$ are the secondary fields of those in the region
 $(\!(1)\!)$.  In general, the fields in the region $(\!(n+1)\!)$
 or $(\!(n+1)\!)'$ are the secondaries
 of the fields in $(\!(n)\!)$ or $(\!(n)\!)'$.
   Since the secondary fields correspond to null vectors,
 those fields  decouple. 
  One can thus construct consistent conformal field theory
  which include only the primary fields in the region $(\!(1)\!)$
 (i.e. inside the minimal table), that is,
  the $(p,q)$ minimal model \cite{BPZ}.
    Coupled to Liouville theory, however, the fields outside the
  the minimal table fail to decouple \cite{Kitazawa} and 
 infinite physical states emerge accordingly.
 These states are considered to correspond to
 the primaries outside 
 the minimal table.
   This correspondence is implied by the
 BRST cohomology  \cite{LZ-BMP} of the coupled system.

  In the minimal model coupled to Liouville theory,
 the gravitational dimension 
$\Delta^{G}_{r,s}$
 of the dressed operator for $\Phi_{r,s}$
 is given by the relations
\beq
\Delta^{G}_{r,s}=1-\frac{\alpha_{r,s}}{\gamma}
\;,
\quad
 \frac{\alpha_{r,s}}{\gamma}
 =\frac{p+q-|pr-qs|}{2q}
 \; ,
\label{eq:alpha-rs}
\eeq
 where $r$ and $s$ take the values in the range \eq{eq:rs-range}.
    On the other hand, in the matrix model, the corresponding relation 
  for the scaling operator $\hsig{j}$ is given by
\beq
 \frac{\alpha_{j}}{\gamma}
 =\frac{p+q-j}{2q}
 \;\; .
\label{eq:alpha-j}
\eeq
 From \eq{eq:alpha-rs} and \eq{eq:alpha-j}, we should take as
\beq
 j=|pr-qs| \;\; , \;\; j=1,2,\cdots\;\ne 0\;(\mbox{mod}\;q)
 \;\; ,
\label{eq:j=}
\eeq
 for $\hsig{j}$.

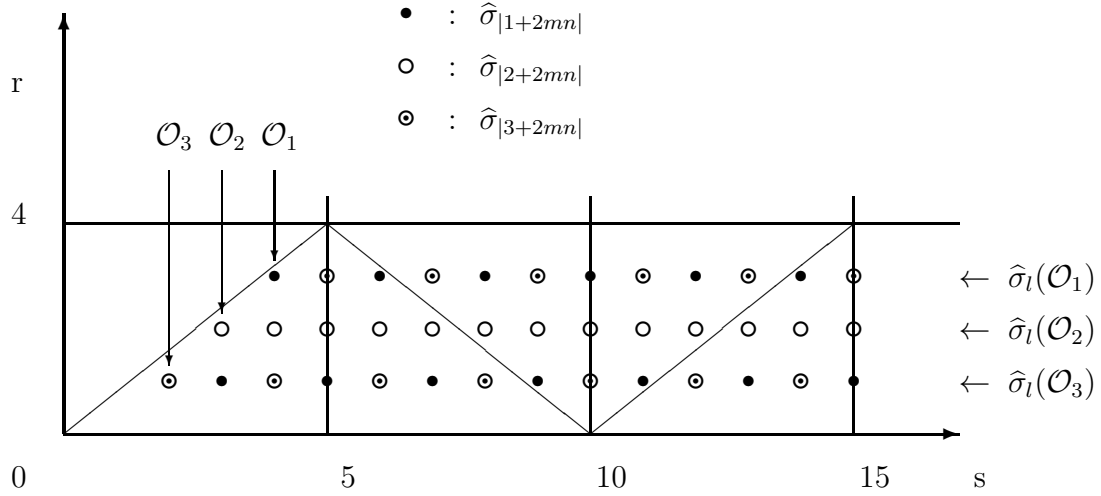
\begin{figure}
 \begin{center}
\setlength{\unitlength}{0.7mm}
\begin{picture}(300,85)(-10,-10)
 \put(0,0){\thicklines\vector(1,0){170} }
 \put(0,0){\thicklines\vector(0,1){80} }
 \put(0,40){\thicklines\line(1,0){170} }
 \multiput(50,0)(50,0){3}{\thicklines\line(0,1){45} }
 \multiput(0,0)(100,0){2}{\line(5,4){50} }
 \put(50,40){\line(5,-4){50} }
 \multiput(40,30)(20,0){6}{\circle*{2}}
 \multiput(30,10)(20,0){7}{\circle*{2}}
 \multiput(30,20)(10,0){13}{\thicklines\circle {2.5}}
 \multiput(50,30)(20,0){6}{\thicklines\circle {2.5}}
 \multiput(50,30)(20,0){6}{\circle*{1}}
 \multiput(20,10)(20,0){7}{\thicklines\circle {2.5}}
 \multiput(20,10)(20,0){7}{\circle*{1}}
 \put(-10,-10){\makebox(8,8)[bl]{0}}
 \put(-10,65){\makebox(8,8)[bl]{r}}
 \put(-10,40){\makebox(8,8)[bl]{4}}
 \put(170,-10){\makebox(8,8)[b]{s}}
 \put(50,-10){\makebox(8,8)[b]{5}}
 \put(100,-10){\makebox(8,8)[b]{10}}
 \put(150,-10){\makebox(8,8)[b]{15}}
 \put(170,28){\makebox(10,30)[bl]
    {$\gets\;\hsig{l}\bigl({\cal O}_1\bigr)$} } 
 \put(170,18){\makebox(10,30)[bl]
    {$\gets\;\hsig{l}\bigl({\cal O}_2\bigr)$} } 
 \put(170,8){\makebox(10,30)[bl]
    {$\gets\;\hsig{l}\bigl({\cal O}_3\bigr)$} } 
  \put(20,50){\vector(0,-1){37}}
  \put(30,50){\vector(0,-1){27}}
  \put(40,50){\vector(0,-1){17}}
  \put(16,55){\makebox(10,10)[b]
    {${\cal O}_3$} } 
  \put(26,55){\makebox(10,10)[b]
    {${\cal O}_2$} } 
  \put(36,55){\makebox(10,10)[b]
    {${\cal O}_1$} } 
 \put(65,80){\circle*{2}}
 \put(73,78){\makebox(10,30)[bl]
    {: $\;\hsig{|1+2mn|}$} } 
 \put(65,70){\thicklines\circle{2.5}}
 \put(73,68){\makebox(10,30)[bl]
    {: $\;\hsig{|2+2mn|}$} } 
 \put(65,60){\circle*{1}}
 \put(65,60){\thicklines\circle{2.5}}
 \put(73,58){\makebox(10,30)[bl]
    {: $\;\hsig{|3+2mn|}$} } 
\end{picture}
 \end{center}
\caption{the scaling operators $\hsig{|k+2mn|}$
 in the $(5,4)$ minimal model coupled to 2d gravity}
\label{fig:pic2}
\end{figure}

   Consider now  the relation of $\hsig{|k+2mn|}$
 to the primary field $\Phi_{r,s}$ of the unitary $(m+1,m)$
 minimal model.
   Let us first compare the two sets 
\beq
 S_{k}=\Bigl\{|k+2nm|\;\bigm|\;n\in\ZZ \;\Bigr\}
 \;\; ,
\label{eq:Sk}
\eeq
 and
\beqn
 \Bigl\{\;|pr-qs|\;\Bigr\}&=&\Bigl\{\;|(m+1)r-ms|\;\Bigr\}
 \nonumber \\
 &=& \Bigl\{\;r'+(s-r-1)m\;\Bigr\}
 \;\; ,
\label{eq:|pr-qs|}
\eeqn
 where $r$ and $s$ are positive integers in the range
\beq
   1\le r\le m-1 ,\quad
   1\le s  , \quad
   r+1\le s
\label{eq:rs-range-2}
\eeq
 and $r'\equiv m-r$.
 Note that we include $s=0$ (mod $m+1$) here.
 Decomposing the set $S_k$ into  two sets as
\beq
 S_{k}=S^+_{k}\oplus S^-_{k}
 \;\; ,
\label{eq:Sk-2}
\eeq
where
\beqn
 S^+_{k}&=&\Bigl\{k+2nm\;\bigm|\;n=0,1,2,\cdots\;\Bigr\}
 \;\; , \nonumber \\
 S^-_{k}&=&\Bigl\{(m-k)+(2n'+1)m\;\bigm|\;n'=0,1,2,
 \cdots\;\Bigr\}
 \;\; ,
\label{eq:S+S-}
\eeqn
 and comparing \eq{eq:S+S-} and \eq{eq:|pr-qs|},
 we can express  the sets $S^+_k$ and $S^-_k$
 in terms of $|(m+1)r-ms|$ as
\beqn
 \!\!\!
 S^+_k&=&\Bigl\{\;|(m+1)r-ms|\;\bigm|\;r'=k,\;s-r=2n+1,
 \;n=0,1,\cdots\; \Bigr\}
 \; ,
 \nonumber \\
 \!\!\!
 S^-_k&=&\Bigl\{\;|(m+1)r-ms|\;\bigm|\;r=k,\;s-r=2n'+2,
 \;n'=0,1,\cdots\;\Bigr\}
 \; .
\label{eq:S+S--2}
\eeqn
 From \eq{eq:S+S--2}, the following correspondence is obtained:
\beqn
 \hsig{|k+2mn|}\;( n\ge 0) &\leftrightarrow&
 \Phi_{m-k,\;r+2n+1} \; ( n\ge 0)
 \nonumber \\
 \hsig{|k+2m(-1-n')|}\;( n'\ge 0 ) &\leftrightarrow&
 \Phi_{k,\;r+2n'+2} \; ( n'\ge 0)
 \;\;,
\label{eq:sigPhi}
\eeqn
 where $s\ne 0\;(\mbox{mod}\;m+1)$.
 As for the scaling operators corresponding to $s=0$ (mod $m+1$),
 we will discuss these in the next section.

   As an example, we depicted the scaling operators
 on the r-s plane for the case of $m=4$ in 
 fig.~\ref{fig:pic2}.
  In this figure we showed the equivalence classes
 $[\hsig{k}]$ explicitly.

\begin{center}
\section{Boundary operators}
\end{center}

\subsection{Boundary operators and touching of loops}
\hspace{5mm}
 The scaling operators $\hsig{j}$ ($j=0$ mod $m+1$), which
 correspond to $s=0$ (mod $m+1$) on the r-s plane do not
 have their counterparts in the BRST cohomology of the
 minimal model coupled to Liouville theory.
  In \cite{MMS} it was proposed that the scaling operators
 which do not occur in the BRST cohomology of Liouville
 theory are boundary operators and  one of them,
 which is $\hsig{3}=\hsig{1}({\cal O}_1)$ in the case of
 pure gravity, was in fact proven to be a boundary operator
 for the one-matrix model and the Ising model case.
   We would like to examine the role played by
 the operators $\hsig{n(m+1)},\;n=1,2,\cdots\ne 0\;
 ({\rm mod}\;m)$ 
 as well as $\hsig{m+1}$ for general unitary minimal models.

\begin{figure}
\begin{center}
 \epsfxsize=12cm
 \qquad\epsfbox{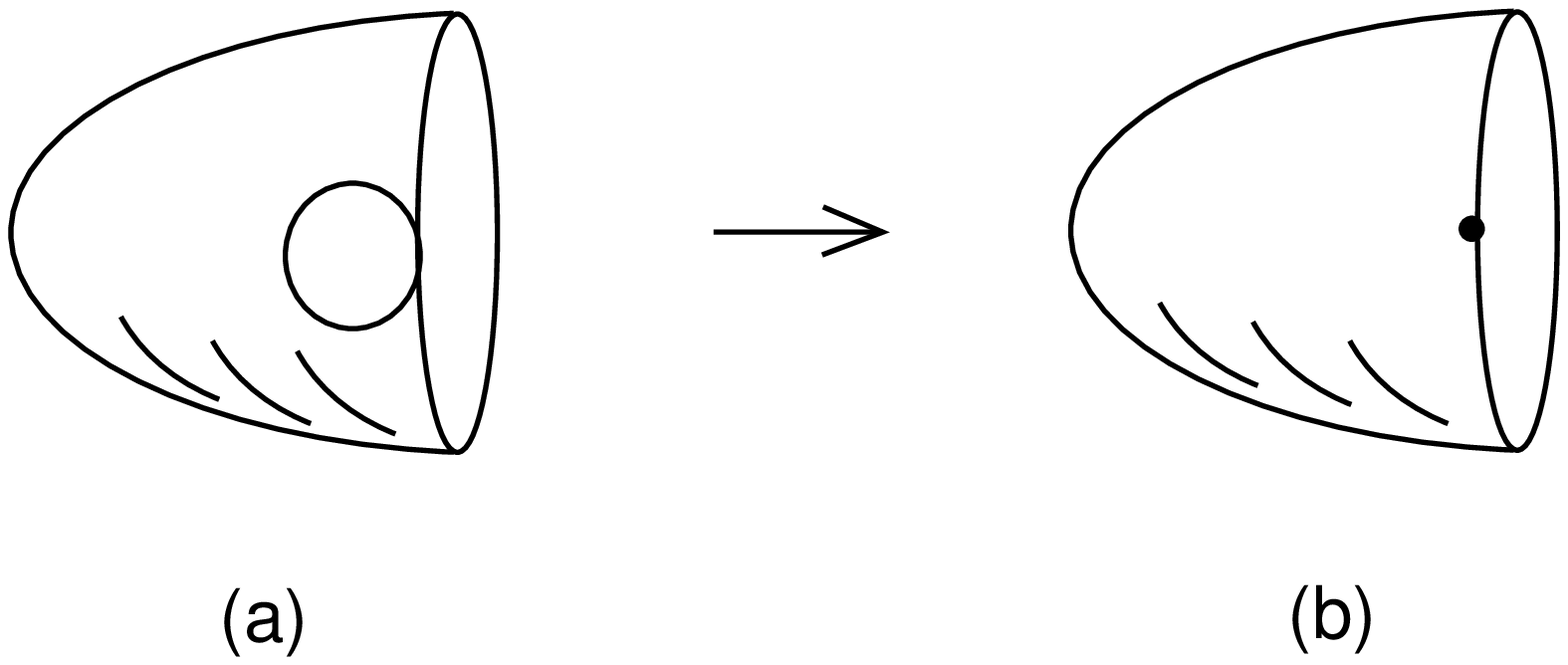}
\caption{(a): A surface with two loops touching each other 
 on a point.
 (b): When one of the loops shrinks to a microscopic loop
 the microscopic loop is equivalent to the insertion of the 
 operator denoted by the dot on the loop.}
\label{fig:bound1}
\end{center}
\begin{center}
 \epsfxsize=12cm
 \qquad\epsfbox{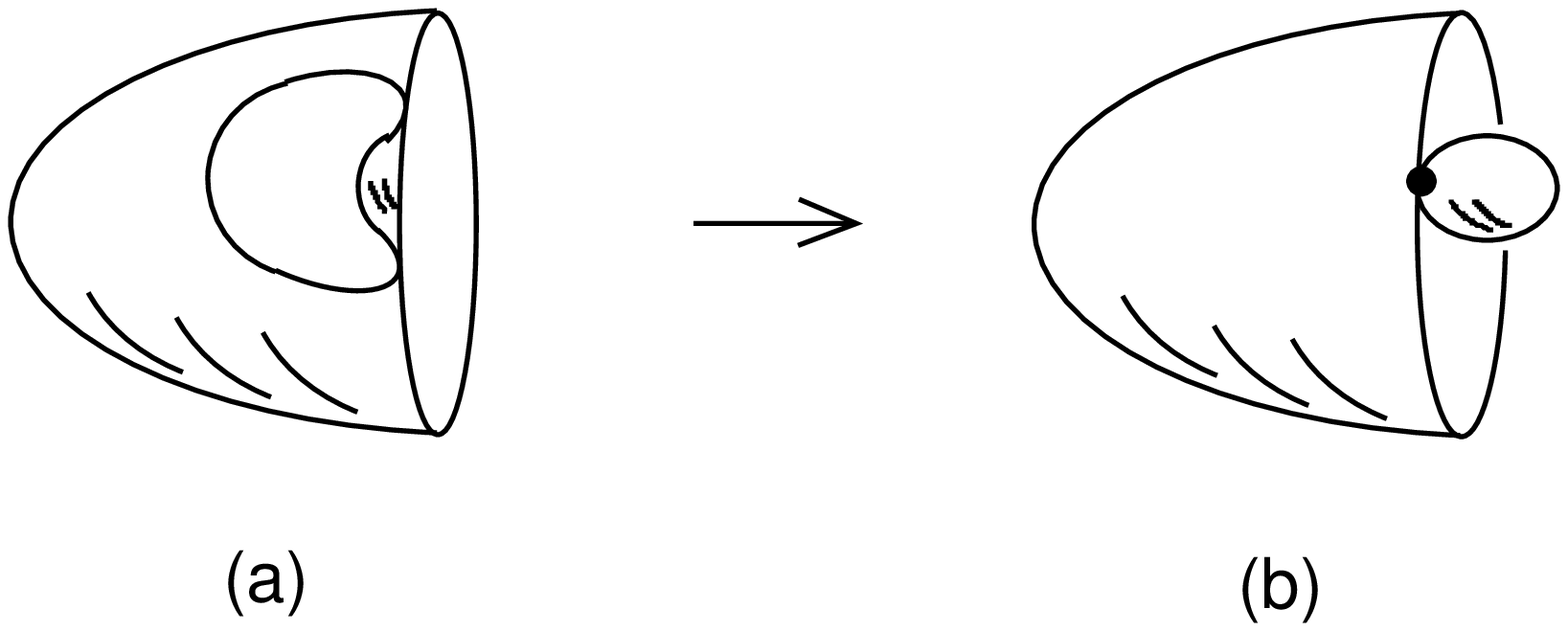}
\caption{The case of a surface with two loops touching each other
 on two points.}
\label{fig:bound2}
\end{center}
\begin{center}
 \epsfxsize=12cm
 \qquad\epsfbox{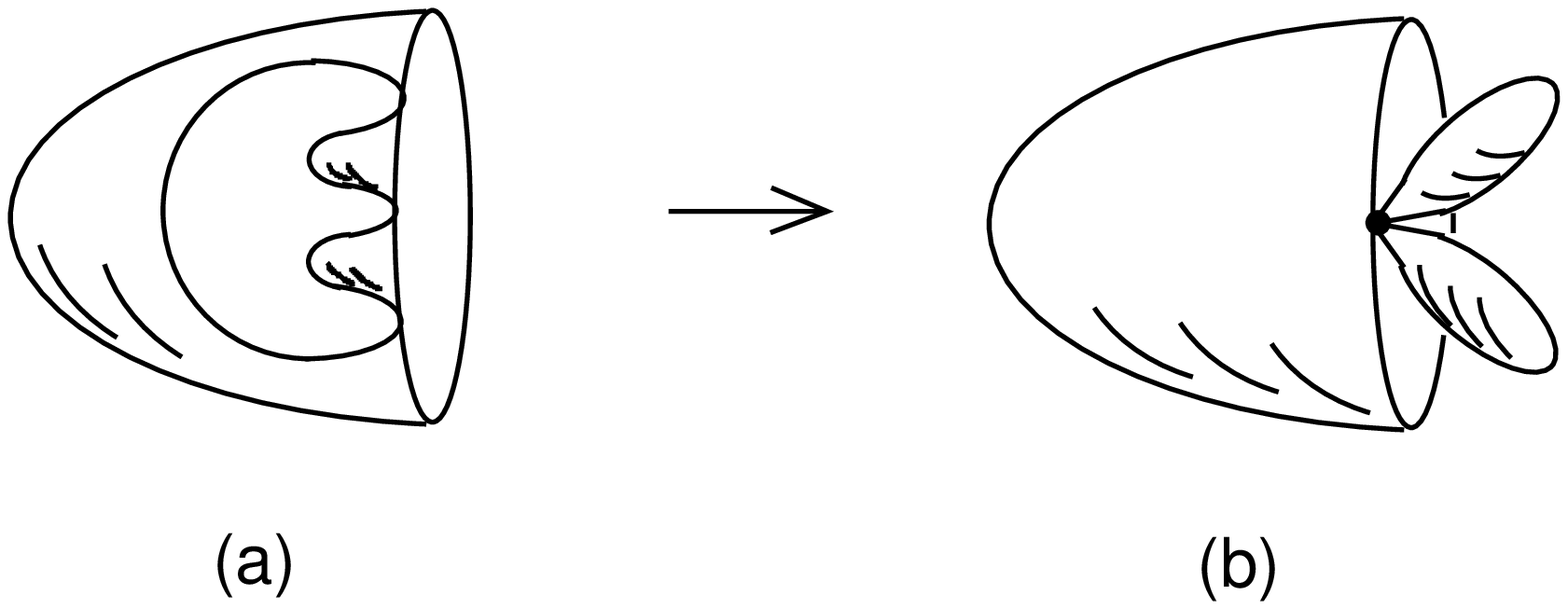}
\caption{The case of a surface with two loops touching each other
 on three points.}
\label{fig:bound3}
\end{center}
\end{figure}

  Let us denote these operators by
\beq
 \widehat{{\cal B}}_n=\hsig{n(m+1)}
 ,\;\; n=1,2,\cdots\ne 0\; ({\rm mod}\;m)
 \;.
\eeq
  In the matrix models the loop amplitudes contain the
 contribution from the surfaces with loops touching
 each other.
  In two-loop case, let us consider the surfaces
 in which the two loops touch  each other on n points.
  When we shrink one of the loops to a microscopic loop,
 the other loop splits into
 n loops, which are stuck each other through the
 microscopic loop 
 (see figs.\ref{fig:bound1}, \ref{fig:bound2}
 and \ref{fig:bound3}).
  Since the microscopic loop represents
 a sum of the scaling operators, the wave functions of
 some scaling operators contain the contribution from the
 surfaces with split loop.
   
  We now show that the boundary operators indeed
 represent these surfaces.
  From eqs. (\ref{eq:wave-function+}) and (\ref{eq:w+-}),
 the wave function of $\widehat{{\cal B}}_n$
 and the one-loop amplitude are
\beq
 \VEV{\widehat{{\cal B}}_n w^+(\l)}
 =n(1+\frac 1m) \left(\frac M2\right)^{n(1+\frac 1m)}
 \tK{n(1+\frac 1m)}(M\l)
\;,
\eeq
\beq
 \VEV{w^+(\l)} =(1+\frac 1m) ~\l^{-1}
 \left(\frac M2\right)^{1+\frac 1m}
 \tK{1+\frac 1m}(M\l)
\;.
\eeq
 We denote the Laplace transformation of 
 any function $f(\l_1,\l_2,\cdots)$ of loop lengths by
\beq
 {\cal L}\left[f(\l_1,\l_2,\cdots)\right]
 =\int_{0}^{\infty}d\l_1 \int_{0}^{\infty}d\l_2 \cdots
 e^{-(\l_1 \zeta_1+\l_2 \zeta_2+\cdots)}
 f(\l_1,\l_2,\cdots)
\;.
\eeq
 In the space of Laplace transformed coordinate $\zeta$,
 we have
\beq
 {\cal L}\left[\l^{-1}
 \langle \widehat{{\cal B}}_n w^+(\l)\rangle\right]
 =-\left(\frac M2\right)^{n(1+\frac 1m)}
 2 \cosh n(m+1)\theta
\;,
\eeq
\beq
 {\cal L}\left[ \langle w^+(\l)\rangle \right]
 =-\left(\frac M2\right)^{1+\frac 1m}
 2 \cosh (m+1)\theta
\;,
\eeq
 where we have used the relation
\beq
 {\cal L} \left[-\l^{-1} |\nu| \tK{\nu}(M\l)
 \right]
 = 2\cosh m\nu\theta
\;,
\eeq
 and $\zeta$ is parametrized as
$
 \zeta=M\cosh m\theta
$.
  Note here that $w^+(\l)$ represents a loop with a marked
 point and $\l^{-1}w^+(\l)$ represents a loop without a
 marked point.
  Since $\cosh n(m+1)\theta$ can be expressed as
 a polynomial of $\cosh (m+1)\theta$,
\beqn
 &&2\cosh n(m+1)\theta
 =2~ T_n\Bigl(\cosh (m+1)\theta\Bigr)
 \nonumber \\
 &&\qquad=
 \sum_{r=0}^{[n/2]} c^{(n)}_{r}
 \Bigl[2\cosh (m+1)\theta\Bigr]^{n-2r}
 ,\quad 
 c^{(n)}_r=\frac{(-1)^r n}{n-r} \left(
 {n-r \atop r}\right)
\eeqn
 where $T_n$ is the Chebeyshev polynomial,
 we obtain the following relation:
\beq
 {\cal L}\left[ -\l^{-1}\langle \widehat{{\cal B}}_n
 w^+(\l)\rangle \right]
 = \sum_{r=0}^{[(n-1)/2]} c^{(n)}_{r}
 \left(\frac M2\right)^{2r(1+\frac 1m)}
 \left\{ {\cal L}\left[-\langle w^+(\l)\rangle \right]
 \right\}^{n-2r}
\;.
\eeq
  In the space of loop length, the above relation means
 that the wave function of $\widehat{{\cal B}}_n$
 is equivalent to a sum of the convolutions of disk amplitudes:
\beq
\label{eq:bound-1}
 \VEV{\widehat{{\cal B}}_n w^+(\l)}
 =-\l \sum_{r=0}^{[(n-1)/2]}c^{(n)}_{r}
 \left(\frac M2\right)^{2r(1+\frac 1m)}
 (-1)^{n-2r}\left[\odot {\cal A}_1^+\right]^{n-2r}(\l)
\;.
\eeq
 Here we introduced a notation
  ${\cal A}_1^+\equiv \langle w^+(\l)\rangle$,
 and $\left[\odot {\cal A}_1^+\right]^{s}(\l)$ denotes
 the convolution of $s$ ${\cal A}_1^+(\l)$'s, for example
\beq
 \left[\odot {\cal A}_1^+\right]^{2}(\l)
  =\int_0^{\infty}\int_0^{\infty} \d\l_1\d\l_2
 \;\delta(\l_1+\l_2-\l) 
 {\cal A}_1^+(\l_1){\cal A}_1^+(\l_2)
\;.
\eeq
   From \eq{eq:bound-1} we can conclude that
 the operator ${\widehat{\cal B}}_n$ couple to the point
 to which $s$ ($ \le n$) parts of the loop are stuck
 each other in the case of one-loop amplitudes.
   When there are more than one loop, we infer
 that the operator couples to the point to which
 $s$ parts of several loops are stuck each other;
 the operator will not recognize that it is
 touching different loops this time.

 Using the following relation
\beq
\label{eq:cosh x}
 \Bigl[2\cosh x\Bigr]^{n}
 =\sum_{r=0}^{[(n-1)/2]} \left({n\atop r}\right)
 2 \cosh (n-2r)x 
,\;\;(\mbox{up to constant})
\;,
\eeq
 we also obtain
\beq
\label{eq:bound-2}
 \l\left[\odot{\cal A}_1^+\right]^{n}(\l)
 =(-1)^{n+1}\sum_{r=0}^{[(n-1)/2]} \left({n\atop r}\right)
 \left(\frac M2\right)^{2r(1+\frac 1m)}
 \VEV{\widehat{\cal B}_{n-2r} w^+(\l)}
\;.
\eeq
  Here we drop the constant term 
 in \eq{eq:cosh x} when
 we carry out the inverse Laplace transformation.
 From \eq{eq:bound-2}, we see that the boundary operator
 coupled to the point on which n parts of loops
 are touching each other  is given by
\beq
 {\cal B}_n
 =(-1)^{n+1}\sum_{r=0}^{[(n-1)/2]} \left({n\atop r}\right)
 \left(\frac M2\right)^{2r(1+\frac 1m)}
 \widehat{\cal B}_{n-2r}
\;.
\eeq

 Now let us consider the boundary operators when
 there are two loops on two-dimensional surface.
   As for ${{\cal B}}_1$, we expect that 
  $ \VEV{w^+(\l_1)w^+(\l_2){{\cal B}}_1}$
  should be proportional to
  $(\l_1+\l_2)\VEV{w^+(\l_1)w^+(\l_2)}$.
  Let us confirm this in the following.
 From the three loop correlator (\ref{eq:w+w+w+-2}),
 the expansion of loop operator (\ref{eq:w+--exp}) and
 the wave function of $\hsig{|k+2mn|}$
 (\ref{eq:wave-function+}),
 we obtain the following correlator with two loops and
 a local operator:
\beqn
\label{eq:2loop+op}
 &&\VEV{w^+(\l_1)w^{\pm}(\l_2)\hsig{|k_3+2mn_3|}}
 =\frac{-1}{m+1}\sum_{k_1,k_2}C_{k_1 k_2 k_3}
 (\pm)^{k_2-1}
 \left(\frac M2\right)^{-\frac 1m +|\frac {k_3}m+2n_3|}
 \nonumber \\
 &&\qquad\qquad\qquad\times
 \l_1\l_2(\frac {k_3}m +2n_3)
 \tK{1-\frac{k_1}m}(M\l_1)\tK{1-\frac{k_2}m}(M\l_2)
\;.
\eeqn
  Consider the amplitude for
 ${{\cal B}}_1=\widehat{{\cal B}}_1=\hsig{m+1}=\hsig{|\frac{m-1}m-2|}$.
 Since $C_{k_1,k_2,m-1}$ is nonvanishing only for
 the case of $k_1+k_2=m$,
 we have
\beq\label{eq:2loop-bound}
 \VEV{w^+(\l_1)w^{\pm}(\l_2){{\cal B}}_1}
 =\frac{1}{m}\sum_{k}^{m-1}
 (\pm)^{k-1}
 \left(\frac M2\right) \l_1\l_2
 \tK{\frac{k}m}(M\l_1)\tK{1-\frac{k}m}(M\l_2)
\;.
 \eeq
 Comparing eqs. (\ref{eq:2loop-bound}) to (\ref{eq:w+w+-})
 we obtain the desired relation
\beq
 \VEV{w^+(\l_1)w^{\pm}(\l_2){{\cal B}}_1}
 =\left\{\l_1+(-1)^m \l_2\right\} \VEV{w^+(\l_1)w^{\pm}(\l_2)}
\;.
\eeq
 Note here that we have
$
 \VEV{{\cal B}_1 w^+(\l)}
 =(-1)^m  \VEV{ {\cal B}_1 w^-(\l)}
$.
 
 Next, let us consider ${\cal B}_2$.
 Since we infer that the insertion of ${\cal B}_2$
 should  play the role of connecting two parts
 of loops together, we expect the relation
\beqn
\label{eq:2loop+B2}
  &&\VEV{w^+(\l_1)w^+(\l_2){\cal B}_2}
 =2\l_1\int_0^{\l_1}d\l'_1 ~
 \VEV{w^+(\l'_1)w^+(\l_2)}\VEV{w^+(\l_1-\l'_1)}
 \nonumber \\
 &&\qquad\qquad
 +2\l_2\int_0^{\l_2}d\l'_2 ~
 \VEV{w^+(\l_1)w^+(\l'_2)}\VEV{w^+(\l_2-\l'_2)}
 \nonumber \\
 &&\qquad\qquad
 +2\l_1 \l_2 \VEV{w^+(\l_1+\l_2)}
 \;.
\eeqn
  The third term in the right hand side of \eq{eq:2loop+B2}
 represents the contribution from the surfaces with
 loops $w^+(\l_1)$ and  $w^+(\l_2)$ touching
  with each other on a point.
 Let us confirm the relation (\ref{eq:2loop+B2})
 in the following.
 In this case, it is convenient to consider in the 
 space of Laplace transformed coordinates $\zeta_i$.
 In this space \eq{eq:2loop+op} reads as
\beqn
 &&\VEV{\hat{W}^+(\zeta_1)\hat{W}^{\pm}(\zeta_2)
 ~\hsig{|k_3+2mn_3|} }
 =\frac{-1}{m+1}
 \left(\frac M2\right)^{-\frac 1m -2 +|\frac {k_3}m+2n_3|}
 (\frac {k_3}m +2n_3)
 \nonumber \\
 &&\quad\times
 \frac{\p}{\p\zeta_1}\frac{\p}{\p\zeta_2}
 \left\{
 \sum_{k_1,k_2}C_{k_1 k_2 k_3}
 (\pm)^{k_2-1}
 ~\frac{\sinh (m-k_1)\theta_1}{\sinh m \theta_1}
 ~\frac{\sinh (m-k_2)\theta_2}{\sinh m \theta_2}
 \right\}
\;,
\eeqn
 where $\hat{W}^{\pm}(\zeta_i)\equiv{\cal L}[w^{\pm}(\l_i)]$.
 Due to the relation
\beqn
 && \sum_{k_1,k_2}C_{k_1 k_2 k_3}
 (\pm)^{k2}
 ~\frac{\sinh (m-k_1)\theta_1}{\sinh m \theta_1}
 ~\frac{\sinh (m-k_2)\theta_2}{\sinh m \theta_2}
\nonumber \\
 &&\quad
 =\frac{-1}{2(\cosh\theta_1 \mp \cosh\theta_2)}
 \left(
 \frac{\sinh (m-k_3)\theta_1}{\sinh m \theta_1}
 -(\pm)^{k_3} \frac{\sinh (m-k_3)\theta_2}{\sinh m \theta_2}
 \right)
 ,
\eeqn
 we have
\beqn
 &&\VEV{\hat{W}^+(\zeta_1)\hat{W}^{\pm}(\zeta_2)
 ~\hsig{|k_3+2mn_3|} }
 =\frac{\pm 1}{2(m+1)}
 \left(\frac M2\right)^{-\frac 1m -2 +|\frac {k_3}m+2n_3|}
 (\frac {k_3}m +2n_3)
 \nonumber \\
 &&\quad\times
 \frac{\p}{\p\zeta_1}\frac{\p}{\p\zeta_2}
 \left\{
 \frac{1}{\cosh\theta_1 \mp \cosh\theta_2}
 \left(
 \frac{\sinh (m-k_3)\theta_1}{\sinh m \theta_1}
 -(\pm)^{k_3} \frac{\sinh (m-k_3)\theta_2}{\sinh m \theta_2}
 \right)
 \right\}
\;.
\nonumber \\
\eeqn
 Since we should take $k_3=2$ for
 ${\cal B}_2=-\hsig{2(m+1)}$ (for $m\ge 3$),
 we obtain the amplitude for ${\cal B}_2$
\beqn
\label{eq:2loop+B2:2}
 &&\VEV{\hat{W}^+(\zeta_1)\hat{W}^{\pm}(\zeta_2)
 {\cal B}_2 }
 \nonumber \\
 &&\;
 =\frac{\mp 1}{m}
 \left(\frac M2\right)^{\frac 1m}
 \frac{\p}{\p\zeta_1}\frac{\p}{\p\zeta_2}
 \left\{
 \frac{1}{\cosh\theta_1 \mp \cosh\theta_2}
 \left(
 \frac{\sinh (m-2)\theta_1}{\sinh m \theta_1}
 -\frac{\sinh (m-2)\theta_2}{\sinh m \theta_2}
 \right)
 \right\}
 .
\nonumber \\
\eeqn
  On the other hand, from the amplitudes \cite{DKK}
\beqn
 &&\VEV{\hat{W}^+(\zeta_1)\hat{W}^+(\zeta_2)}
 = \frac{\p}{\p\zeta_1}\frac{\p}{\p\zeta_2}
 \ln \frac{\cosh\theta_1-\cosh\theta_2}
 {\cosh m\theta_1-\cosh m\theta_2}
 \nonumber \\
 &&\qquad\qquad
 =\frac{\p}{\p \zeta_2}
 \left\{
 \frac{1}{\cosh\theta_1-\cosh\theta_2}
 \frac{\sinh\theta_1}{mM\sinh m\theta_1}
 -\frac{1}{\zeta_1-\zeta_2}
 \right\}
\;,
\eeqn
 we obtain the relation
\beqn
\label{eq:2loop+B2:3}
 &&-2\frac{\p}{\p\zeta_1}
 \left\{
 \VEV{\hat{W}^+(\zeta_1)\hat{W}^+(\zeta_2)}
 \VEV{\hat{W}^+(\zeta_1)}
 \right\}
 +(1\leftrightarrow 2)
\nonumber \\
 &&-2\frac{\p}{\p\zeta_1}\frac{\p}{\p\zeta_2}
 \left\{
 \frac{\VEV{\hat{W}^+(\zeta_1)}-\VEV{\hat{W}^+(\zeta_2)} }
 {\zeta_1-\zeta_2}
 \right\}
\nonumber \\
 &&\;
 =\frac{2}{m}
 \left(\frac M2\right)^{\frac 1m}
 \frac{\p}{\p\zeta_1}\frac{\p}{\p\zeta_2}
 \left\{
 \frac{1}{\cosh\theta_1-\cosh\theta_2}
 \left(
 \frac{\sinh\theta_1 \cosh (m+1)\theta_1}{\sinh m \theta_1}
 -(1\leftrightarrow 2)
 \right)
 \right\}
 .
\nonumber \\
\eeqn
 One can easily show  that the right hand side of
 eq.~(\ref{eq:2loop+B2:3}) agrees with that of 
 eq.~(\ref{eq:2loop+B2:2}). 
 Putting eqs.~(\ref{eq:2loop+B2:3}) and
 (\ref{eq:2loop+B2:2}) together and performing 
 the inverse Laplace transformation, we obtain the desired
 relation eq.~(\ref{eq:2loop+B2}).

 As for $\VEV{w^+(\l_1)w^-(\l_2){\cal B}_2 }$,
 from the amplitude \cite{DKK}
\beq
 \VEV{\hat{W}^+(\zeta_1)\hat{W}^-(\zeta_2)}
 = \frac{\p}{\p\zeta_1}\frac{\p}{\p\zeta_2}
 \ln (\cosh\theta_1+\cosh\theta_2)
\;,
\eeq
 we obtain the relation
\beqn
\label{eq:2loop+B2:4}
  &&\VEV{w^+(\l_1)w^-(\l_2){{\cal B}}_2}
 =-2\l_1\int_0^{\l_1}d\l'_1 ~
 \VEV{w^+(\l'_1)w^-(\l_2)}\VEV{w^+(\l_1-\l'_1)}
 \nonumber \\
 &&\qquad\qquad
 +2\l_2\int_0^{\l_2}d\l'_2 ~
 \VEV{w^+(\l_1)w^-(\l'_2)}\VEV{w^-(\l_2-\l'_2)}
 \;
\eeqn
  in a similar way.
 In this case, the operator ${\cal B}_2$ does not connect
 the different kinds of loops $w^+(\l_1)$ and $w^-(\l_2)$
 together.

  We have shown that the operator ${\cal B}_2$ connects
 two parts of the same kind of loops  together
 in the case with two loops.
  We infer that similar phenomena occur in general;
 the operator ${\cal B}_n$ 
 connects n parts of the same kind of loops together
 in the case with any number of loops.
 
\subsection{Connection to the Schwinger-Dyson equations}
\hspace{5mm}
  We can observe a close relationship between  the boundary
 operators and the Schwinger-Dyson equations proposed in
 \cite{IIKMNS}.
  Continuum limit of the 
 Schwinger-Dyson equations for loops in the two-
 and multi-matrix models were proposed in \cite{IIKMNS}
 under some assumptions. It was shown \cite{IIKMNS} that
 these equations for the two-matrix model
 contain $W_3$ constraints,
 which were derived explicitly in \cite{GN}.
  The integrability of these equations were shown in
 \cite{S-D eqs.}.
  These facts justify the proposed Schwinger-Dyson
 equations.

 Let us consider the connection of the boundary
 operators with the Schwinger-Dyson equations.
  For the $(m+1,m)$ minimal models, the following
 Schwinger-Dyson equations were proposed in
 \cite{IIKMNS}:
\beqn
\label{eq:S-D}
 &&\int_0^{\l} d\l'
 \VEV{w^{(1)}(\l')
 w^{(1)}\left(\l-\l';[{\cal H}(\sigma)]^j \right)
 w^{(1)}(\l_1)\cdots w^{(1)}(\l_n) }'
 \nonumber \\
 &&
 +g \sum_i \l_i
 \VEV{
 w^{(1)}\left(\l+\l_i;[{\cal H}(\sigma)]^j \right)
 w^{(1)}(\l_1)\cdots w^{(1)}(\l_{i-1})
 w^{(1)}(\l_{i+1})\cdots w^{(1)}(\l_n)
 }'
 \nonumber \\
 &&
 +\VEV{
 w^{(1)}\left(\l;[{\cal H}(\sigma)]^{j+1} \right)
 w^{(1)}(\l_1)\cdots w^{(1)}(\l_n)
 }'
 \approx 0 \;,
 \nonumber \\
 &&\qquad\qquad\qquad
 {\rm for}\;\; j=0,\cdots,m-2
\;,
\eeqn
 and
\beq
\label{eq:S-D:2}
 \VEV{
 w^{(1)}\left(\l;[{\cal H}(\sigma)]^{m-1} \right)
 w^{(1)}(\l_1)\cdots w^{(1)}(\l_n)
 }'
 \approx 0
 \;.
\eeq

 Here $\VEV{\cdots}'$ represent loop correlators
 that are not necessarily connected,
 and $w^{(1)}(\l)$ represents a loop operator
 corresponding to a loop created by the matrix
 $\hat{A}^{(1)}$
 in the multi-matrix model.
  The operator ${\cal H}(\sigma)$ describes an operator
 which changes  the `spin' on a loop locally from 1 to 2.
 Also $w^{(1)}\left(\l;[{\cal H}(\sigma)]^{j} \right)$
 describes a loop with $[{\cal H}(\sigma)]^{j}$ inserted.
 The symbol $\approx$ means that as a function of $\l$,
 the quantity has its support at $\l=0$.

 From \eq{eq:S-D} for $j=0$ and $n=1$, we have the relation
\beqn
 &&\l_1\VEV{ w^{(1)}\left(\l_1;{\cal H}(\sigma) \right)
 w^{(1)}(\l_2)}'
 +\l_2\VEV{ w^{(1)}(\l_1)
 w^{(1)}\left(\l_2;{\cal H}(\sigma) \right)}'
\nonumber \\
 &&
 +\l_1\int_0^{\l_1}d\l'_1
 \VEV{ w^{(1)}(\l'_1)w^{(1)}(\l_1-\l'_1)w^{(1)}(\l_2)}'
\nonumber \\
 &&
 +\l_2\int_0^{\l_2}d\l'_2
 \VEV{ w^{(1)}(\l_1)w^{(1)}(\l'_2)w^{(1)}(\l_2-\l'_2)}'
\nonumber \\
 &&
 +2 g\l_1\l_2  \VEV{ w^{(1)}(\l_1+\l_2)}'
 \approx 0
\; .
\label{eq:S-D:3}
\eeqn
 The planar part of the above relation agrees with
 \eq{eq:2loop+B2}. Note that the loop amplitudes in
 \eq{eq:2loop+B2} represent connected correlators.

  This agreement implies that ${\cal H}$ would correspond
 to $\widehat{{\cal B}}_2$.
 Taking into account the fact that $\widehat{{\cal B}}_n$
 ($n=0$ mod $m$) do not exist and \eq{eq:S-D:2},
 it is legitimate to consider that the amplitude
 (for $j=1,\cdots,m$)
\beq
 \VEV{w^+(\l_1)\cdots w^+(\l_n)\widehat{{\cal B}}_j}
\eeq
 corresponds to the connected part of the amplitude
\beq
 \sum_{i=1}^{n}
 \oint d\sigma_i
 \VEV{
  w^{(1)}(\l_1)\cdots w^{(1)}(\l_{i-1})
 w^{(1)}\left(\l_i;[{\cal H}(\sigma_i)]^{j-1} \right)
 w^{(1)}(\l_{i+1})\cdots w^{(1)}(\l_n)
 }'
\;.
\nonumber \\
\eeq
\begin{center}
\section{Summary}
\label{summary}
\end{center}

   In this paper we have examined the correlators
 in the $(m+1,m)$ unitary minimal models coupled to
 two-dimensional gravity from the point of view of
 the two-matrix model.
 From the two-loop correlators and the wave function of the
 scaling operators, we derived the explicit form of
 the expansion of the loops
 in terms of the scaling operators.
  Using this expansion, we deduced the three-point functions
 from the three loop operators, and
 showed that simple fusion rules exist for all of the
 scaling operators.
  The three-loop correlator \cite{AII2} can be understood
 to express these fusion rules in a compact form.

  At the $(m+1,m)$ critical point in two-matrix models,
 the scaling operators
 $\hsig{j}$ $(j=0$  mod $m+1)$ have no counterparts
 in the BRST cohomology of Liouville theory coupled
 to the corresponding conformal matter.
   In \cite{MMS},these operators were argued to be
 boundary operators
 which couple to loops in the case of the one-matrix
 model. It was also shown explicitly that one of them,
 corresponding to $\hsig{m+1}$ in the case of 
 the unitary matter, is a 
 operator which measures the total length of the loops.

    We examined the role played by the rest of these operators.
 We showed, in some examples,
 that the operator ${\cal B}_n$ couples to
 the points to which n parts of the same kind of loops
 are stuck
 each other. In other words, the operator ${\cal B}_n$
 connects n parts of the same kind of loops  together.
  We think these operators
 play an important role concerning the touching of the
 macroscopic loops. The emergence of these operators
 in matrix models can easily be understood from the viewpoint
 of macroscopic loops and their expansion in terms of local
 operators.

\begin{center}
\section{Acknowledgements}
\label{ack}
\end{center}
 
 We thank Hiroshi Itoyama for helpful discussions and the
 careful reading of the manuscript.
  We also thank Keiji Kikkawa for useful discussions on
 this subject.

\appendix
\begin{center}
\section{}
\end{center}

 Let us prove that the two-loop correlator
  \eq{eq:w+w+-} can be written as \eq{eq:w+w+--exp}.
  From \eq{eq:w+w+-}, we have
\beq
 \VEV{w^+(\l_1) w^{\pm}(\l_2)}
 =\sum_{k=1}^{m-1} (\pm)^{k-1}
 \left( \frac{\sin \pi \frac km}{\pi/2} \right)^{2} \;
 \VEV{w_{k}(\l_1) w_{k}(\l_2)}
\label{eq:w+w+-sum}
\eeq
and 
\beq
 \frac{\p}{\p M}
 \VEV{w_{k}(\l_1) w_{k}(\l_2)}
 =
 -\frac 1{m} \frac M2 \l_1 \l_2 \; 
 K_{1-\frac km}(M\l_1)\;K_{1-\frac km}(M\l_2)
 \; ,
\label{eq:Mwkwk}
\eeq
 where we have introduced loop operators $w_k(l)$ which
 represent loops with  some distinct matter boundary
 condition.
   Making use of a formula 
\beq
 K_{\nu}(z)\;K_{\nu}(w)
 =\frac 12 \int_{0}^{\infty} \frac{\d t}{t}\;
 K_{\nu}\left(\frac{zw}{t}\right)\;
 \exp\left(-\frac t2 -\frac{z^2+w^2}{2t}\right)
\label{eq:KnuzKnuw}
\eeq
 and replacing $t$ with $tM^2$, we have
\beqn
 \lefteqn{ 
 M \l_1 \l_2 \; K_{1-\frac km}(M\l_1)\;
 K_{1-\frac km}(M\l_2)
 } \nonumber \\
 &&=
 \frac 12 \int_{0}^{\infty} \frac{\d t}{t}\;
 K_{1-\frac km} \left(\frac{\l_1\l_2}{t}\right)\;
 \exp\left(-\frac {tM^2}{2} -\frac{\l_1^2+\l_2^2}{2t}\right)
 \;\; .
\label{eq:int-1}
\eeqn
  Carrying out the integral with respect to $M$, and from
 \eq{eq:Mwkwk}, we have
\beqn
 \lefteqn{
 \VEV{w_{k}(\l_1) w_{k}(\l_2)}
 } \nonumber \\
 &&
 = \frac 1{4m} 
 \int_{0}^{\infty} \frac{\d t}{t}\;
 \frac{\l_1\l_2}{t}
 K_{1-\frac km} \left(\frac{\l_1\l_2}{t}\right)\;
 \exp\left(-\frac {tM^2}{2} -\frac{\l_1^2+\l_2^2}{2t}\right)
 \; .
\label{eq:int-2}
\eeqn
 Due to a formula,
\beq
 z K_{1-|p|}(z)
 =
 \int_{0}^{\infty}\d E
 \frac{E\sinh \pi E}{\cosh \pi E -\cos \pi p}
 \; K_{iE}(z)
 \;\; , 
\label{eq:zK-int}
\eeq
 the right hand side of \eq{eq:int-2} turns into
\beqn
\lefteqn{
 \frac 1{4m}
 \int_{0}^{\infty} \frac{\d t}{t}
 } \nonumber \\
 &&\times
 \int_{0}^{\infty}\d E
 \frac{E\sinh \pi E}{\cosh \pi E -\cos \pi p}
 \; K_{iE} \left( \frac{\l_1 \l_2}{t} \right)
 \exp\left(-\frac {tM^2}{2} -\frac{\l_1^2+\l_2^2}{2t}\right)
 .
\label{eq:int-3}
\eeqn
 Using a formula \eq{eq:KnuzKnuw} again, \eq{eq:int-3} turns out to be
\beq
 \frac 1{2m}
 \int_{0}^{\infty}\d E
 \frac{E\sinh \pi E}{\cosh \pi E -\cos \pi \frac km}
 \; K_{iE}(M\l_1) \; K_{iE}(M\l_2)
 \; .
\label{eq:int-4}
\eeq
 Putting \eq{eq:int-4} and \eq{eq:w+w+-sum} together,
 we have
\beqn
 \lefteqn{
 \VEV{w^+(\l_1) w^{\pm}(\l_2)}
 =
 \sum_{k=1}^{m-1} \frac 1{2m} (\pm)^{k-1}
 \left( \frac{\sin \pi \frac km}{\pi/2} \right)^{2} \;
 } \nonumber \\ 
 && \qquad\qquad\times
 \int_{0}^{\infty}\d E
 \frac{E\sinh \pi E}{\cosh \pi E -\cos \pi \frac km}
 \; K_{iE}(M\l_1) \; K_{iE}(M\l_2)
\label{eq:w+w+-off}
\eeqn

 The integral in $E$ can be carried out  by
 deforming the contour.
 The residues for poles
$
 E=\pm i (\frac km +2n)
 , \; n=0,\pm 1, \pm 2, \cdots
 \; ,
$
 contribute to the integral and, after all, we obtain the
 following expansion for the two-loop correlators (for
 $\l_1 < \l_2$) 
\beqn
 \lefteqn{
 \VEV{w^+(\l_1) w^{\pm}(\l_2)}
 } \nonumber \\
 &&=
 \frac 1m \sum_{k=1}^{m-1} \sum_{n=-\infty}^{\infty} \;
 (\pm)^{k-1}
 \Bigl|\frac km +2n\Bigr|\; I_{|\frac km +2n|}(M\l_1)
 \; \tK{\frac km +2n}(M\l_2)
 \; .
\label{eq:w+w+--exp:app}
\eeqn

\vspace{1cm}
\begin{center}

\end{center}
\end{document}